\documentclass[journal]{IEEEtran}

\usepackage{graphicx}
\usepackage{amsmath,epsfig,amssymb}
\begin{document}
\title{New Power Estimation Methods for Highly Overloaded Synchronous CDMA Systems}

\author{D.~Nashtaali, O.~Mashayekhi, \textit{Student Member}, IEEE, P.~Pad, \textit{Student Member}, IEEE, S.~R.~Moghadasi, and F.~Marvasti, \textit{Senior Member}, IEEE \thanks{D.~Nashtaali, O.~Mashayekhi, P.~Pad, and F.~Marvasti are with Advanced Communications Research Institute (ACRI) and Department of Electrical Engineering, Sharif University of Technology, Tehran, Iran (\{d\_nashtaali,omashayekhi,pedram\_pad\}@ee.sharif.edu and marvasti@sharif.edu).}\thanks{S.~R.~Moghadasi is with Department of Mathematical Sciences at Sharif University of Technology, Tehran, Iran (moghadasi@sharif.edu).}}\maketitle

\begin{abstract}
In CDMA systems, the received user powers vary due to moving distance of users. Thus, the CDMA receivers consist of two stages. The first stage is the power estimator and the second one is a Multi-User Detector (MUD). Conventional methods for estimating the user powers are suitable for under- or fully-loaded cases (when the number of users is less than or equal to the spreading gain). These methods fail to work for over-loaded CDMA systems because of high interference among the users. Since the bandwidth is becoming more and more valuable, it is worth considering overloaded CDMA systems. In this paper, an optimum user power estimation for over-loaded CDMA systems with Gaussian inputs is proposed. We also introduce a suboptimum method with lower complexity whose performance is very close to the optimum one. We shall show that the proposed methods work for highly over-loaded systems (up to $m\left(m+1\right)/2$ users for a system with only $m$ chips). The performance of the proposed methods is demonstrated by simulations. In addition, a class of signature sets is proposed that seems to be optimum from a power estimation point of view. Additionally, an iterative estimation for binary input CDMA systems is proposed which works more accurately than the optimal Gaussian input method.
\end{abstract}


\section{Introduction}\label{sec:Intro}
\PARstart{I}{n} a CDMA system, every user is assigned a signature vector that is used for transmitting data through a common channel. In the channel, all the transmitted vectors are added up and the noisy sum reaches the receiver end.

Because of the variable distance of users from the base station, the received powers of the users differ from their desired values. This phenomenon is called the near-far effect; power control techniques are used to combat this effect. However, because these techniques are imperfect, the receiver needs to estimate the received power for each user for proper decoding. This problem becomes more critical when the system works in an overloaded environment in which the number of users is greater than the spreading gain. Thus, the general structure for a CDMA receiver consists of two parts, the first one is a power estimation module and the second one is an MUD module.

So far, for combating the near-far effect in CDMA, efforts have been devoted to finding near-far resistance detectors. MUD rather than single user detection can significantly improve the performance \cite{verdu}. However, because optimum detector is very complex actually NP complete, suboptimum detectors are considered. Some of these suboptimum detectors are called linear and constrained optimization methods \cite{xli}-\cite{jtug}. Subspace-based signature waveform estimation by using Wiener \cite{umad}, \cite{umad2} and Kalman \cite{xzhan} are other approaches. Another method is based on parametric signal modelling and signal spectrum estimation \cite{xia}-\cite{xwan}.   
Some other strategies for combating the near-far effects are the covariance method in combination with successive interference cancellation \cite{ycho}, blind adaptive interference suppression \cite{umad3} and isolation bit insertion \cite{fc}. Additionally, upper and lower bounds for near-far resistance of a covariance detector is found in \cite{umad4}. Another approach to mitigate the near-far effect is the channel estimation at the receiver. Adaptive filter techniques have been used to estimate the channel in \cite{dek}. However, in overloaded systems, the performances of these methods degrade due to high multiuser interference. 

Recently, some studies have been performed for CDMA systems that do not need a power estimation unit at the receiver. The discussion in \cite{Hossein} shows that the performance of such systems depends not only on the receiver structure but also on the signature sequences that are allocated to each user. Signature matrices that guarantee errorless communication in overloaded systems, when the channel has near-far effects in the absence of noise, are studied in the same paper. In \cite{sumcap}, assuming that the near-far effects of the users have a Gaussian distribution, the authors have determined some upper and lower bounds for the sum channel capacity of the large scale binary CDMA systems.

Nonetheless, the model that has been used for the channels in \cite{Hossein} and \cite{sumcap} are much worse than what occurs in practice. They assume that the stochastic process of the power change of each user is a white process, and the received value of the power of each user is independent of its value in any other time index. Yet, in practical situations, the fact is that the user powers change slowly (in comparison with the data rate); thus we can assume that it has a piecewise-constant behavior.

This paper proposes the Maximum Liklihood (ML) estimation for the user powers even for very highly overloaded CDMA systems when the user data and the channel noise have Gaussian distribution which achives the channel capacity. The method achieves the exact values of the powers asymptotically. The only information that is used by the receiver is the signatures of the users and the covariance matrix of the channel noise. It will be shown that given a CDMA system with chip rate $m$ and no more than $m\left(m+1\right)/2$ users, there exists a signature matrix for which the ML estimation has unique solution and yields perfect power estimation asymptotically. For the systems that have more than $m\left(m+1\right)/2$ users, the ML estimation has no longer a unique solution. 

In addition, two suboptimum estimators are proposed. The first one is a suboptimum version of the proposed ML estimation that has lower computational cost with similar performance. Additionally, this study will introduce a characteristic about how a signature matrix is good from power estimation perspective and a new class of codes will be proposed that seems to be optimum with respect to this characteristic.

The second method is an iterative power estimation in case of binary input vectors. This iterative method works much more accurately with just a few number of received vectors. It will be proved that for a noiseless channel with user powers constrained to be in a predefined interval, this method can find the powers exactly. Simulation results show that for high noise variance and wide range of powers, the estimator remains accurate.

The performances of the proposed methods are simulated for different situations when the channel noise variance is known or unknown at the receiver. The ability of tracking the changes of the powers is also simulated. Moreover, a CDMA system is simulated that uses binary signatures in conjunction with BPSK modulation utilizing the proposed methods for power estimation and is also compared to the system with perfect or no power control. The simualtion results show excellent performances for these methods.

In the next section, the channel model is presented. In Section \ref{sec:ML-Decoder}, the ML power estimation is introduced. In Section \ref{sec:method}, a suboptimum estimation is proposed. We also introduced a class of accurate signature matrices for estimating user powers. The iterative estimation is discussed in section \ref{sec:IterBinary}. These two methods are simulated in a practical system using a BPSK modulation in section \ref{sec:simulation}. Conclusions and future works are in section \ref{sec:conclusion}.

\section{Channel Model}\label{sec:Prelim}
In a synchronous CDMA system, each user is assigned an $m$-chip signature vector for transmitting its data through the channel. Each user multiplies its data by its signature vector before transmission. The channel noise is added to the transmitted vectors with different powers at the receiver. Assuming a synchronous CDMA system with $n$ users such that the received power of the $i^{th}$ user is $p_i$, $i=1,\dots,n$, the channel can be modeled as    
\begin{eqnarray}\label{equ:Linmodel}
Y=\mathbf{S}\mathbf{P}^{1/2}X+N
\end{eqnarray}
where $\mathbf{S}$ is the $m\times n$ signature matrix in which the $i^{th}$ column is the signature of user $i$, $\mathbf{P}=\text{diag}\left(p_1,\dots,p_n\right)$, $X$ is the user data vector, $N$ is the channel noise vector, and $Y$ is the received vector. Notice that it is assumed that the columns of the signature matrix $\mathbf{S}$ are normalized and the entries of $X$ have unity power.

Because of the variable distance of the users from the base station, the values of the user powers $p_i$'s change and thus the matrix $\mathbf{P}$ is unknown at the receiver. Therefore, at the receiver we need to estimate the user powers for proper extracting data $X$ from the received vector $Y$.

In practical situations, the data rate is high enough such that we can reasonably assume that $\mathbf{P}$ is constant over a number of received vectors, say $L$. In other words, the variation of the matrix $\mathbf{P}$ versus time is piecewise constant.

Power estimation is not difficult in under- or fully-loaded systems ($n\leq m$) since orthogonal signatures that provide an independent channel for each user can be used. However, in over-loaded systems ($n>m$), where such codes do not exist, the problem becomes much more complicated. Nevertheless, over-loaded systems are preferred because of their bandwidth efficiencies. The following section proposes the ML estimation of the matrix $\mathbf{P}$ for a system with any given number of users with Gaussian distribution which achives the channel capacity.

\section{ML Power Estimation}\label{sec:ML-Decoder}
In this section using (\ref{equ:Linmodel}), we wish to find the maximum likelihood estimation of the user powers matrix, $\mathbf{P}=\text{diag}\left(p_{1},\ldots,p_{n}\right)$, from the observed vectors $Y_1,\ldots,Y_L$. We assume that in these $L$ observations the user powers are constant. Thus, we have
\begin{eqnarray}\label{equ:MLRelation}
\hat{\mathbf{P}}_{ML}=\arg\max_{\mathbf{P}}{~f\left(Y_{1},\ldots,Y_{L}\vline~\mathbf{P}\right)}
\end{eqnarray}
where $\hat{\mathbf{P}}_{ML}$ is the ML-estimation of the user powers matrix and $f$ is the Probability Density Function (PDF) of the observed vectors. Hereafter in this section, we assume that the data vector has Gaussian distribution with covariance matrix $\mathbb{E}\{XX^{T}\}=\mathbf{I}_{m}$ and the noise vector $N$ has distribution $\mathcal{N}\left(\mu,\mathbf{\Sigma}\right)$.

Since $Y_i$'s for $i=1,\ldots,L$ are $i.i.d.$ Gaussian random vectors with mean $\mu$ and covariance matrix of $\mathbf{M_P}=\mathbf{SPS}^T+\mathbf{\Sigma}$, we have
\begin{align}\label{equ:MLIndep}
\hat{\mathbf{P}}_{ML}=&\arg\max_{\mathbf{P}}{~\prod_{i=1}^{L} f\left(Y_{i}\vline~\mathbf{P}\right)}=\arg\max_{\mathbf{P}}{~\ln\left(\prod_{i=1}^{L} f\left(Y_{i}\vline~\mathbf{P}\right)\right)}\nonumber\\=&\arg\max_{\mathbf{P}}{~\sum_{i=1}^{L}\ln{f\left(Y_{i}\vline ~\mathbf{P}\right)}}\nonumber\\=&\arg\max_{\mathbf{P}}~\sum_{i=1}^{L}\bigg(-\frac{1}{2}\left(Y_{i}-\mu\right)^{T} \mathbf{M_P}^{-1}\left(Y_{i}-\mu\right)\nonumber\\&-\ln{\sqrt{\left(2\pi\right)^{m} \left|\det\left(\mathbf{M_P}\right)\right|}}\bigg)
\end{align}

Thus, we deduce that,
\begin{align}\label{equ:MLRelationforDiff}
\hat{\mathbf{P}}_{ML}=\arg\min_{\mathbf{P}}\bigg(&\frac{1}{L}\sum_{i=1}^{L}\left(Y_{i}-\mu\right)^{T}\mathbf{M_P}^{-1} \left(Y_{i}-\mu\right)\nonumber\\&+\ln{\left|\det\left(\mathbf{M_P}\right)\right|}\bigg)
\end{align}

Notice that the estimation of the user powers is possible if the solution of the above equation is unique. The following theorem expresses conditions under which the above minimization has a unique solution.
\\

\newtheorem{theo}{\textbf{Theorem}}
\begin{theo}\label{th:MLTOTAL}
Let $\bar{\bar{\mathbf{S}}}$ be an $\frac{m\left(m+1\right)}{2}\times n$ matrix whose rows are the entry-by-entry multiplication of the rows of the signature matrix $\mathbf{S}$. Then, the minimization (\ref{equ:MLRelationforDiff}) has a unique solution if and only if 
\begin{eqnarray}
\text{rank}\left(\bar{\bar{\mathbf{S}}}\right)=n
\end{eqnarray}
\end{theo}

We prove the necessary condition of Theorem \ref{th:MLTOTAL}; and the converse will be proved for sufficiently large $L$'s.

\emph{Proof of necessary condition}: Suppose that $n>\text{rank}\left(\bar{\bar{\mathbf{S}}}\right)$. Then, there exists an $n\times 1$ non-zero vector $\underline{B}$ such that $\bar{\bar{\mathbf{S}}}\underline{B}=0$. It yields that $\mathbf{SB}\mathbf{S}^T=\mathbf{0}$ where $\mathbf{B}$ is an $n\times n$ diagonal matrix that has the entries of $\underline{B}$ as its diagonal entries. Now, since for any diagonal matrix $\mathbf{P}$, $\mathbf{M_P}=\mathbf{S}\mathbf{P}\mathbf{S}^{T}+\mathbf{\Sigma}=\mathbf{S} \left(\mathbf{P+B}\right)\mathbf{S}^{T}+\mathbf{\Sigma}=\mathbf{M_{P+B}}$, the minimization (\ref{equ:MLRelationforDiff}) has no unique solution.

\emph{Proof of the converse theorem}: We desire to show that for sufficiently large $L$, the second derivative of (\ref{equ:MLRelationforDiff}) is positive definite and thus it is a convex function and has a unique minimum. Let
\begin{align}\label{equ:Funcs}
&\Phi\left(\mathbf{P}\right)=\frac{1}{L}\sum_{i=1}^{L}\left(Y_{i}-\mu\right)^{T}\mathbf{M}^{-1}_\mathbf{P}\left(Y_{i}-\mu\right)+\ln{\left|\det\left(\mathbf{M_P}\right)\right|}
\end{align}
Then we have,
\begin{align}\label{equ:DiffNormMat}
\Psi\left(\mathbf{A},\mathbf{P}\right)&=D_{\mathbf{P}}\Phi\left(\mathbf{A}\right)=\text{tr}\left(\mathbf{M}^{-T}_\mathbf{P}\mathbf{S}\mathbf{A}\mathbf{S}^{T}\right)
\nonumber\\&-\frac{1}{L}\sum_{i=1}^{L}\left(Y_{i}-\mu\right)^{T}\mathbf{M}^{-1}_\mathbf{P} \left(\mathbf{S}\mathbf{A}\mathbf{S}^{T}\right)\mathbf{M}^{-1}_\mathbf{P}\left(Y_{i}-\mu\right)\nonumber\\
&=\text{tr}\left(\mathbf{M}^{-1}_\mathbf{P}\mathbf{S}\mathbf{A}\mathbf{S}^{T}\right)\nonumber\\&-\text{tr}\left(\frac{1}{L}\sum_{i=1}^{L}\left(Y_{i}-\mu\right)^{T}\mathbf{M}^{-1}_\mathbf{P}\left(\mathbf{S}\mathbf{A}\mathbf{S}^{T} \right)\mathbf{M}^{-1}_\mathbf{P}\left(Y_{i}-\mu\right)\right)\nonumber\\&=
\text{tr}\left(\mathbf{S}^{T}\mathbf{M}^{-1}_\mathbf{P}\mathbf{S}\mathbf{A}\right)-\nonumber\\&\text{tr}\left(\mathbf{S}^{T} \mathbf{M}^{-1}_\mathbf{P}\left(\frac{1}{L}\sum_{i=1}^{L}\left(Y_{i}-\mu\right)\left(Y_{i}-\mu\right)^{T}\right) \mathbf{M}^{-1}_\mathbf{P}\mathbf{S}\mathbf{A}\right)\nonumber\\
&=\text{tr}\left(\mathbf{S}^{T}\mathbf{M}^{-1}_\mathbf{P}\left(\mathbf{I}-\mathbf{W}\mathbf{M}^{-1}_\mathbf{P}\right)\mathbf{S}\mathbf{A}\right)
\end{align}
where $D_{\mathbf{P}}\Phi\left(\cdot\right)$ denotes the derivative of $\Phi$ at point $\mathbf{P}$, $\mathbf{W}$ is the estimation of the covariance matrix $\mathbf{M_P}$ from the received vectors, i.e. $\mathbf{W}=\frac{1}{L}\sum^{L}_{i=1}\left(Y_{i}-\mu\right)\left(Y_{i}-\mu\right)^{T}$.
Getting another derivative, we obtain 
\begin{align}\label{equ:2ndDeriveML}
&D_{\mathbf{P}}\Psi\left(\mathbf{A},\mathbf{B}\right)=\nonumber\\ &\text{tr}\left(\mathbf{S}^{T} \mathbf{M}^{-1}_{\mathbf{P}}\mathbf{SA}\mathbf{S}^{T}\mathbf{M}^{-1}_{\mathbf{P}} \left(2\mathbf{W}-\mathbf{M}_{\mathbf{P}}\right)\mathbf{M}^{-1}_{\mathbf{P}}\mathbf{S}\mathbf{B}\right)
\end{align}
where $\mathbf{A}$ and $\mathbf{B}$ are diagonal matrices. For the proof of convexity, we wish to show that $D_{\mathbf{P}}\Psi\left(\mathbf{A},\mathbf{A}\right)\geq 0$ and equality holds only for $\mathbf{A}=\mathbf{0}$. Note that
\begin{align}\label{equ:ConvexML}
&D_{\mathbf{P}}\Psi\left(\mathbf{A},\mathbf{A}\right)=\text{tr}\left(\mathbf{M}^{-1}_{\mathbf{P}}\tilde{\mathbf{A}}\mathbf{M}^{-1}_{\mathbf{P}} \left(2\mathbf{W}-\mathbf{M}_{\mathbf{P}}\right)\mathbf{M}^{-1}_{\mathbf{P}}\tilde{\mathbf{A}}\right)
\end{align}
where $\tilde{\mathbf{A}}=\mathbf{SA}\mathbf{S}^{T}$. We have
\begin{eqnarray}\label{equ:C}
\text{tr}\left(\mathbf{M}^{-1}_{\mathbf{P}}\tilde{\mathbf{A}}\mathbf{M}^{-1}_{\mathbf{P}} \left(2\mathbf{W}-\mathbf{M}_{\mathbf{P}}\right)\mathbf{M}^{-1}_{\mathbf{P}}\tilde{\mathbf{A}}\right)\nonumber\\
=\text{tr}\left(\mathbf{M}^{-1}_{\mathbf{P}}\tilde{\mathbf{A}}\mathbf{M}^{-1}_{\mathbf{P}}\tilde{\mathbf{A}}\mathbf{M}^{-1}_{\mathbf{P}} \left(2\mathbf{W}-\mathbf{M}_{\mathbf{P}}\right)\right)
\end{eqnarray}
Since 
\begin{eqnarray}
\mathbf{M}^{-1}_{\mathbf{P}}\tilde{\mathbf{A}}\mathbf{M}^{-1}_{\mathbf{P}}\tilde{\mathbf{A}}\mathbf{M}^{-1}_{\mathbf{P}}
=\mathbf{M}^{-1/2}_{\mathbf{P}}\left(\mathbf{M}^{-1/2}_{\mathbf{P}}\tilde{\mathbf{A}}\mathbf{M}^{-1/2}_{\mathbf{P}}\right)^2 \mathbf{M}^{-1/2}_{\mathbf{P}}
\end{eqnarray}
and $\mathbf{M}^{-1/2}_{\mathbf{P}}\tilde{\mathbf{A}}\mathbf{M}^{-1/2}_{\mathbf{P}}$ is a symmetric matrix, $\mathbf{M}^{-1}_{\mathbf{P}}\tilde{\mathbf{A}}\mathbf{M}^{-1}_{\mathbf{P}}\tilde{\mathbf{A}}\mathbf{M}^{-1}_{\mathbf{P}}$ is a positive-semidefinite matrix for non-zero $\tilde{\mathbf{A}}$. As $L$ tends to infinity, from law of large numbers, $2\mathbf{W}-\mathbf{M}_{\mathbf{P}}$ tends to $\mathbf{M}_{\mathbf{P}}$ which is a positive-definite matrix. Therefore, for large $L$, according to Lemma \ref{lem:TraceLem} described below, $\text{tr}\left(\mathbf{M}^{-1}_{\mathbf{P}}\tilde{\mathbf{A}}\mathbf{M}^{-1}_{\mathbf{P}}\tilde{\mathbf{A}}\mathbf{M}^{-1}_{\mathbf{P}} \left(2\mathbf{W}-\mathbf{M}_\mathbf{P}\right)\right)\geq 0$ and equality holds when $\tilde{\mathbf{A}}=\mathbf{0}$. Since $\text{rank}\left(\bar{\bar{\mathbf{S}}}\right)=n$, $\tilde{\mathbf{A}}=\mathbf{0}$ is equivalent to $\mathbf{A}=\mathbf{0}$ and thus the proof is complete.$\blacksquare$\\

\newtheorem{lem}{\textbf{Lemma}}
\begin{lem}\label{lem:TraceLem}
If $\mathbf{V}$ and $\mathbf{W}$ are positive-semidefinite matrices such that $\mathbf{V}^{1/2}\mathbf{W}\mathbf{V}^{1/2}\neq\mathbf{0}$, then $\text{tr}\left(\mathbf{VW}\right)>0$.
\end{lem}
\emph{Proof}: $\text{tr}\left(\mathbf{VW}\right)=\text{tr}\left(\mathbf{V}^{1/2}\mathbf{W}\mathbf{V}^{1/2}\right)$.
Since $\mathbf{W}$ is a positive-semidefinite matrix, $\mathbf{V}^{1/2}\mathbf{W}\mathbf{V}^{1/2}$ is also a positive-semidefinite matrix. Thus, $\text{tr}\left(\mathbf{VW}\right)=\text{tr}\left(\mathbf{V}^{1/2}\mathbf{W}\mathbf{V}^{1/2}\right)\geq 0$ and equality holds only when $\mathbf{V}^{1/2}\mathbf{W}\mathbf{V}^{1/2}=\mathbf{0}$.$\blacksquare$
\\

Theorem \ref{th:MLTOTAL} shows that for a given signature matrix $\mathbf{S}$, the $\text{rank}\left(\bar{\bar{\mathbf{S}}}\right)$ is an upper bound for the number of users for which the user powers can be estimated from the received vectors. Obviously, $\text{rank}\left(\bar{\bar{\mathbf{S}}}\right)\leq\frac{m\left(m+1\right)}{2}$. In the next sections, it will be shown that for any $m$, there exist $m\times\frac{m\left(m+1\right)}{2}$ signature matrices $\mathbf{S}$ such that its corresponding $\bar{\bar{\mathbf{S}}}$ is full rank. 

By having a system with chip rate $m$, it is possible to estimate the user powers for up to $m\left(m+1\right)/2$ users. It is interesting that if the number of users is beyond this threshold, the user powers cannot be estimated from the received vectors $Y$'s. Also, note that if $n\leq m\left(m+1\right)/2$, this estimation is asymptotically exact, i.e., if the number of received vectors tends to infinity, the estimated powers tend to the exact values with probabiliy $1$. This fact is very significant because it implies that we can estimate user powers for any practical overloaded CDMA systems since the number $m\left(m+1\right)/2$ is way beyond number of users for any practical situations.

In addition, using (\ref{equ:DiffNormMat}) in the proof of Theorem \ref{th:MLTOTAL}, we can implement the ML power estimator. If $\mathbf{P}=\hat{\mathbf{P}}_{ML}$ minimizes (\ref{equ:MLRelationforDiff}), for any diagonal matrix $\mathbf{A}$, we have
\begin{eqnarray}\label{equ:SysofEqu}
\text{tr}\left(\mathbf{S}^{T}\mathbf{M}^{-1}_\mathbf{P}\left(\mathbf{I}-\mathbf{W}\mathbf{M}^{-1}_\mathbf{P}\right)\mathbf{S}\mathbf{A}\right)=0.
\end{eqnarray}
It is equivalent to that all diagonal entries of $\mathbf{S}^{T}\mathbf{M}^{-1}_\mathbf{P}\left(\mathbf{I}-\mathbf{W}\mathbf{M}^{-1}_\mathbf{P}\right)\mathbf{S}$ are zero, i.e.,
\begin{eqnarray}\label{equ:MLdiagzero}
\left[\mathbf{S}^{T}\mathbf{M}^{-1}_\mathbf{P}\left(\mathbf{I}-\mathbf{W}\mathbf{M}^{-1}_\mathbf{P}\right)\mathbf{S}\right]_{ii}=0,
\end{eqnarray}
for $i=1,\dots,n$. We can solve this system of non-linear equations to obtain the ML estimation of the user powers.

\begin{figure}[t]
\centering
\includegraphics[width=9cm]{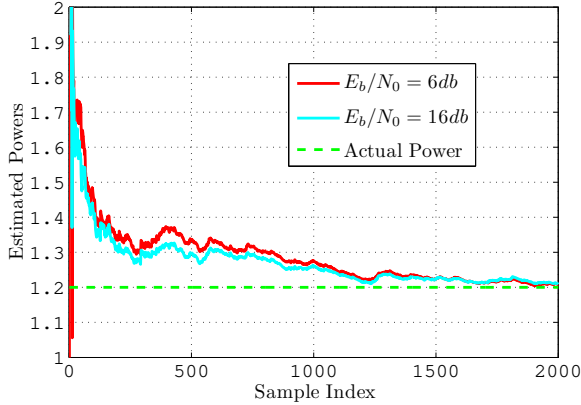}
\caption{The ML estimated power vs. the number of received samples for $E_b/N_0=6\text{dB}$ and $16\text{dB}$.}
\label{fig:F_ML}
\end{figure}

Fig. \ref{fig:F_ML} is the simulation of a system with $12$ users and $8$ chips with binary WBE signatures \cite{KarPad}. We have used the Newton method for solving the above system of equations. The simulation is for AWGN channels with two specific values of $E_b/N_0$.

In the next section, we propose a suboptimum estimator that has a lower computational complexity. Also, we introduce a family of signature matrices which are good from the power estimation point of view.

\section{Suboptimal Power Estimation}\label{sec:method}
In this section, a low complexity method for estimating powers is introduced. Simulation results show that the performance of the proposed low complexity estimator is very close to the ML estimation introduced in the previous section.

\subsection{The Proposed Method}\label{sec:pmethod}
Since $\mathbf{W}$ and $\mathbf{M_P}$ are respectively the empirical and the exact covariance matrices of $Y$, it is expected that $\mathbf{W}$ and $\mathbf{M_P}$ are approximately the same. Notice that, using the law of large numbers, $\mathbf{W}$ tends to $\mathbf{M_P}$ as $L$ tends to infinity. Hence, we try to find the diagonal matrix $\hat{\mathbf{P}}$, such that 
\begin{eqnarray}\label{equ:4}
\mathbf{W}=\mathbf{S}\hat{\mathbf{P}}\mathbf{S}^T+\mathbf{\Sigma}
\end{eqnarray}
for unknowns, $\hat{p}_1,\dots,\hat{p}_n$. We rewrite (\ref{equ:4}) in the form
\begin{eqnarray}\label{equ:star}
\underline{K}=\bar{\bar{\mathbf{S}}}\underline{\hat{P}}
\end{eqnarray}
where $\underline{K}$ is an $\frac{m\left(m+1\right)}{2}\times 1$ vector which contains the entries of the upper triangle of $\mathbf{W}-\mathbf{\Sigma}$, $\bar{\bar{\mathbf{S}}}$ is an $\frac{m\left(m+1\right)}{2}\times n$ matrix defined in Theorem \ref{th:MLTOTAL} and $\underline{\hat{P}}=\left[\hat{p}_1,\dots,\hat{p}_n\right]^T$. In general, we consider the least square solution of (\ref{equ:star}) for $\underline{\hat{P}}$. The same as the result of Theorem \ref{th:MLTOTAL}, the least square solution of (\ref{equ:star}) is unique if and only if the $\text{rank}\left(\bar{\bar{S}}\right)=n$.

\begin{figure}[t]
\centering
\includegraphics[width=9cm]{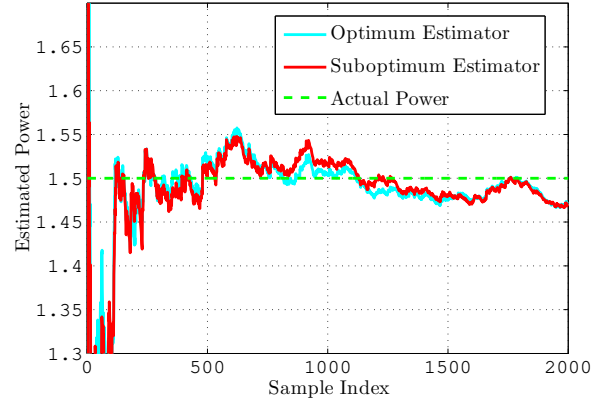}
\caption{The ML and suboptimum estimation of the powers vs. the number of received vectors $L$ for a system with $n=12$ and $m=8$ that uses a binary WBE signature matrix.}
\label{fig:mlcov}
\end{figure}

There is another justification for choosing powers stated in the previous paragraph. In the preceding section, we showed that the ML estimation of the matrix $\mathbf{P}$ should satisfy (\ref{equ:SysofEqu}). Here $\hat{\mathbf{P}}$ is the annihilating term for $\mathbf{I-W}\mathbf{M}_\mathbf{P}^{-1}$ in (\ref{equ:SysofEqu}). Interestingly, simulation results show that the performance of this simpler estimator is as good as the ML one, Fig. \ref{fig:mlcov}.

In practical situations, we face two other problems. Firstly, the covariance matrix of the channel noise is not always known at the receiver end. Secondly, the user powers change by time. Using the suboptimum estimator proposed in this section, we devise solutions for these problems.

If $\mathbf{\Sigma}$ is unknown but the channel noise is white, $\mathbf{\Sigma}$ is a diagonal matrix. In this case, from (\ref{equ:star}) we omit equations that correspond to the diagonal entries of $\mathbf{\Sigma}$. Thus, we arrive at a system of $\frac{m\left(m-1\right)}{2}$ equations,  
\begin{eqnarray}\label{equ:star2}
\underline{\breve{K}}=\breve{\bar{\mathbf{S}}}\underline{\hat{P}}
\end{eqnarray}
where $\breve{\bar{\mathbf{S}}}$ is an $\frac{m\left(m-1\right)}{2}\times n$ matrix wherein each of the rows is the entrywise multiplication of \emph{different} rows of $\mathbf{S}$ (the entrywise multiplication of a row by itself does not appear in $\breve{\bar{\mathbf{S}}})$ and $\underline{\breve{K}}$ is the $\frac{m\left(m-1\right)}{2}\times 1$ vector that contains all entries of $\mathbf{W}$ that are above its main diagonal. Again, in general, we choose the least square solution of (\ref{equ:star2}) as the estimated powers. This solution is unique if and only if the $\text{rank}\left(\breve{\bar{\mathbf{S}}}\right)=n$. It means that for a system with white noises and unknown variances, the maximum number of users cannot be more than $m\left(m-1\right)/2$.

\begin{figure}[t]
\centering
\includegraphics[width=9cm]{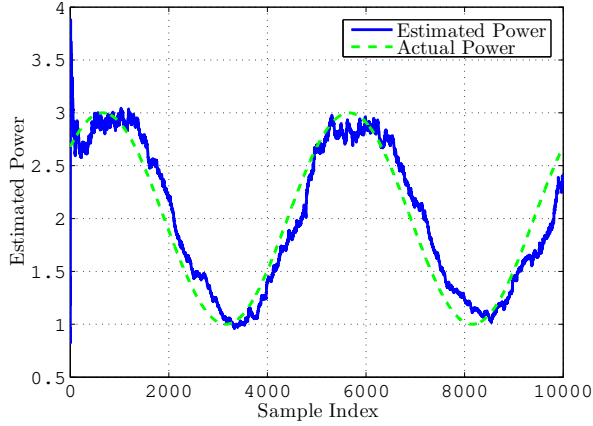}
\caption{The estimated power vs. the number of received samples for the proposed suboptimum estimator for $n=12$, $m=8$ and a binary WBE signature matrix; the user powers are changing with a sinusoidal shape.}
\label{fig:sin}
\end{figure}

\begin{figure}[t]
\centering
\includegraphics[width=9cm]{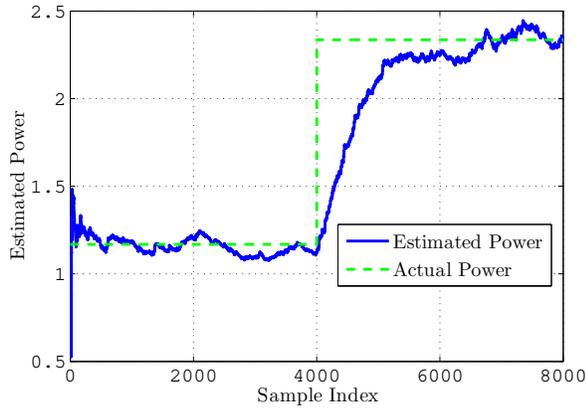}
\caption{The estimated power vs. the number of received samples for the proposed suboptimum estimator for $n=12$, $m=8$ and a binary WBE signature matrix; the user powers is doubled abruptly.}
\label{fig:step}
\end{figure}

To add the ability of tracking the user power variations, we use another window for calculating the empirical covariance matrix of $Y$. By assigning
\begin{eqnarray}\label{eqExp}
\mathbf{W}=\frac{1-\alpha}{1-\alpha^L}\sum_{j=1}^{L}{\alpha^{L-j} Y_jY_j^T}
\end{eqnarray}
where $0\leq\alpha\leq 1$, we decrease the effect of the old samples of the received vectors. Figs. \ref{fig:sin} and \ref{fig:step} show that by this modification, the estimator tracks the power changes rapidly. While it almost perfectly tracks the sinusoidal fluctuation, it is interesting to see that it can also track abrupt changes. 

In the next subsection, we present the signature matrix $\mathbf{S}$ whose corresponding $\bar{\bar{\mathbf{S}}}$ and $\breve{\bar{\mathbf{S}}}$ have rank $n$, respectively. Moreover, the proposed matrices may lead to maximum robustness against errors in $\mathbf{W}$.   

\subsection{Signature Matrix Design for Estimating User Powers}\label{sec:goodmatrices}
It is known that sensitivity of $\underline{\hat{P}}$ to the changes in $\underline{K}$ is equal to the reciprocal of the minimum singular value of $\bar{\bar{\mathbf{S}}}$. Thus, for the proposed estimation method to be more robust against the error in estimating $\underline{K}$, a matrix $\bar{\bar{\mathbf{S}}}$ with larger minimum singular value is needed. More important, if the rank of matrix $\bar{\bar{\mathbf{S}}}$ is less than $n$, its minimum singular value is $0$ and thus (\ref{equ:star}) lacks a unique least square solution. The same statements hold by substituting $\breve{\bar{\mathbf{S}}}$ for $\bar{\bar{\mathbf{S}}}$ and $\underline{\breve{K}}$ for $\underline{K}$ in the system where the noise variance is not known.

In the following theorem, it will be shown that for any $m$ and $n\leq m\left(m+1\right)/2$, there exists $\mathbf{S}_{m\times n}$ such that its corresponding $\bar{\bar{\mathbf{S}}}_{\frac{m\left(m+1\right)}{2}\times n}$ has rank $n$.

\begin{theo}\label{th1}
Suppose $\mathbf{S}_{m\times n}$ is a matrix that its coressponding $\bar{\bar{\mathbf{S}}}_{\frac{m\left(m+1\right)}{2}\times n}$ has rank $n$. Also, for any $k\leq m+1$, let 
\begin{eqnarray}
\mathbf{S}_{\left(m+1\right)\times\left(n+k\right)}=\left[\begin{matrix}\mathbf{1}_{1\times n}&\vline&\mathbf{I}_{k\times k}\\\mathbf{S}_{m\times n}&\vline&\mathbf{0}_{\left(m+1-k\right)\times k}\end{matrix}\right].
\end{eqnarray}
Then $\bar{\bar{\mathbf{S}}}_{\frac{\left(m+1\right)\left(m+2\right)}{2}\times\left(n+k\right)}$ has rank $n+k$.
\end{theo}

\emph{Proof}: Constructing $\bar{\bar{\mathbf{S}}}_{\frac{\left(m+1\right)\left(m+2\right)}{2}\times\left(n+k\right)}$, we find that $k$ linearly independent columns are added to the columns of $\bar{\bar{\mathbf{S}}}_{\frac{m\left(m+1\right)}{2}\times n}$. Using the assumption that $\bar{\bar{\mathbf{S}}}_{\frac{m\left(m+1\right)}{2}\times n}$ has rank $n$, $\bar{\bar{\mathbf{S}}}_{\frac{\left(m+1\right)\left(m+2\right)}{2}\times\left(n+k\right)}$ has rank $n+k$ and the proof is complete.$\blacksquare$
\\

The following theorem shows that for any $m$ and $n\leq m\left(m-1\right)/2$, there exists $\mathbf{S}_{m\times n}$ such that its corresponding $\breve{\bar{\mathbf{S}}}_{\frac{m\left(m-1\right)}{2}\times n}$ has rank $n$.
\\

\begin{theo}\label{th2}
Suppose $\mathbf{S}_{m\times n}$ is a matrix that its coressponding $\breve{\bar{\mathbf{S}}}_{\frac{m\left(m-1\right)}{2}\times n}$ has rank $n$. Also, for any $k\leq m$, let
\begin{eqnarray}
\mathbf{S}_{\left(m+1\right)\times\left(n+k\right)}=\left[\begin{matrix}\begin{matrix}\\\mathbf{0}_{1\times n}\\ \\\mathbf{S}_{m\times n}\\~\end{matrix}&\vline&\begin{matrix}\mathbf{1}_{1\times k}\\ \\\mathbf{I}_{k\times k}\\\\\mathbf{0}_{\left(m-k\right)\times k}\end{matrix}\end{matrix}\right]
\end{eqnarray} 
Then $\breve{\bar{\mathbf{S}}}_{\frac{m\left(m+1\right)}{2}\times\left(n+k\right)}$ has rank $n+k$.
\end{theo}

The proof is simillar to that of Theorem \ref{th1}.
\\

\newtheorem{corol}{\textbf{Corollary}}
\begin{corol}\label{cor1}
Let $\mathbf{S}_{1\times 1}=\left[1\right]$; using Theorem \ref{th1}, we can construct appropriate matrices for power estimation for any $m$ and $n\leq m\left(m+1\right)/2$. Also, using Theorem \ref{th2}, let $\mathbf{S}_{2\times 1}=\left[\begin{matrix}1 & 1\end{matrix}\right]^T$, we can construct suitable matrices for power estimation for any $m$ and $n\leq m\left(m-1\right)/2$.
\end{corol}

Since multiplying each column of $\mathbf{S}$ with a non-zero number leaves the rank of $\bar{\bar{\mathbf{S}}}$ and $\breve{\bar{\mathbf{S}}}$ unchanged, the columns of the matrices introduced in Corollary \ref{cor1} can be normalized to obtain acceptable signature matrices. 

An interesting fact about the normalized version of the matrices constructed in Theorem \ref{th1} is that their $\bar{\bar{\mathbf{S}}}$ have large minimum singular value and thus are almost optimum for estimating the user powers. For the second set of the constructed matrices, in the following theorem we prove that all singular values of the $\breve{\bar{\mathbf{S}}}$ are $0.5$.
\\

\begin{theo}\label{th5}
Let $\mathbf{S}_{m\times n}$ be a signature matrix defined in the Theorem \ref{th2}. All singular values of the $\breve{\bar{\mathbf{S}}}$ are equal to $0.5$.
\end{theo}

\emph{Proof}: From the definition of $\mathbf{S}$, it is deduced that
$\breve{\bar{\mathbf{S}}}$ is formed by concatenating a permutation matrix multiplied with $0.5$ with an all zero matrix. Hence, all the singular values are $0.5$.$\blacksquare$
\\

It is worth mentioning that by checking more than $50000$ normalized random matrices, we did not find any matrices with minimum singular value greater than the ones constructed by Theorems \ref{th1} and \ref{th2}. However, the optimality of these matrices has not been proven yet. 

The performance of the matrices proposed in Theorems \ref{th1} and \ref{th2} are simulated. The simulations were performed for the proposed $8\times 36$ and $8\times 28$ matrices in situations where the noise variance is known and unknown at the receiver end, respectively. Although the overloading factor is very severe, the powers are estimated accurately by a few number of received vectors. Notice that the conventional power estimation methods fail to work even in much lower overloaded systems. 

\begin{figure}[t]
\centering
\includegraphics[width=9cm]{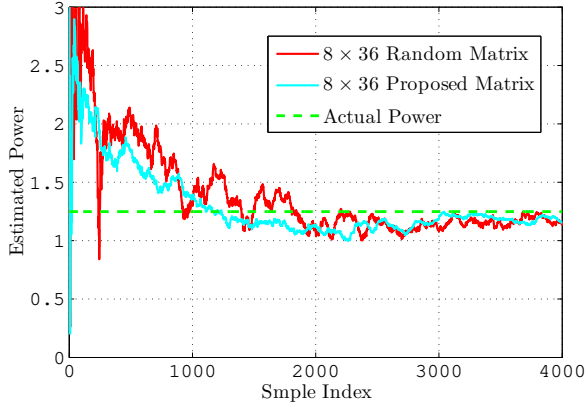}
\caption{The comparison of the estimated power vs. the number of received samples for an $8\times 36$ matrix (Theorem \ref{th1}) and a random matrix of the same size. Here we have an AWGN channel with a known noise variance and $E_b/N_0=6\text{dB}$.}
\label{fig:known}
\end{figure}

\begin{figure}[t]
\centering
\includegraphics[width=9cm]{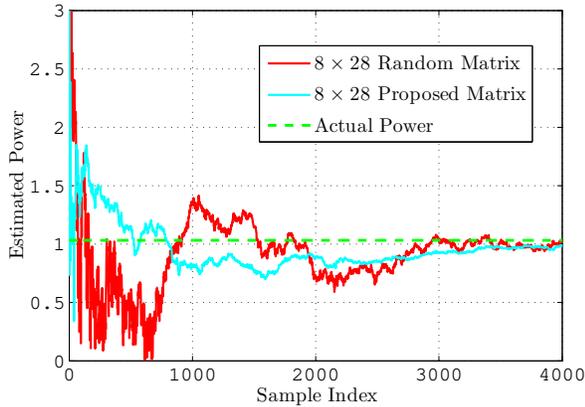}
\caption{The comparison of the estimated power vs. the number of received samples for an $8\times 26$ matrix (Theorem \ref{th2}) and a random matrix of the same size. Here we have an AWGN channel with an unknown noise variance and $E_b/N_0=6\text{dB}$.}
\label{fig:unknown}
\end{figure}

\section{Iterative Power Estimation for Binary Input CDMA System}\label{sec:IterBinary}
In pevious sections, we introduced optimum and suboptimum power estimators in the case of Gaussian input vector. However, in practical situations, binary input systems are more favorable. While estimators proposed in Sections \ref{sec:ML-Decoder} and \ref{sec:method} work for binary input, they are not really optimum. In this section, iterative power estimation for binary input synchronous CDMA systems is developed which is superior to the estimators described previously. Estimations become remarkably better when $E_b/N_0$ is increased.

Indeed it will be proved that with no noise in the channel this method can determine the actual user powers with probability one provided that the variance of the power fluctuations do not exceed a given threshold. However, the simulation results show that this method has an acceptable performance in noisy channels and when power fluctuations are very large.

\begin{figure}[t]
\centering
\includegraphics[width=9cm]{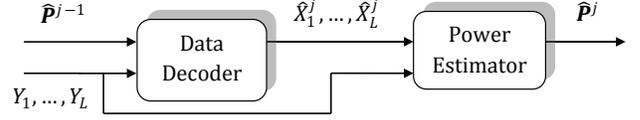}
\caption{The $j^{th}$ iteration of the iterative power estimatior.}
\label{fig:chart}
\end{figure}

We use an iterative scheme for estimating the powers. In each iteration, we use the output power of the previous iteration and decode the user data. Then, using these extracted values of the user data, we again estimate the user powers. Fig. \ref{fig:chart} shows this procedure. We denote the decoded data of the $i^{th}$ received vector and the power matrix in the $j^{th}$ iteration by $\hat{X}_i^j$ and $\hat{\mathbf{P}}^j$, respectively.

According to the supposition that the powers are constant in the interval of sending $L$ vectors, we find the least square solution of the following linear system of $Lm$ equations and $n$ unknowns ($\hat{p}_i$'s for $i=1,\ldots,n$) as the estimation of the user powers.
\begin{eqnarray}\label{equ:i1}
Y_i=\mathbf{S}\sqrt{\hat{\mathbf{P}}^j}\hat{X}_i^{j}~~~~~~~~~~~~i=1,2,\ldots,L
\end{eqnarray}
By Multiplying each column of the signature matrix $\mathbf{S}$ by its corresponding entry of $\hat{X}_i^j$, we get $\hat{\mathbf{S}}_i^j$. Thus (\ref{equ:i1}) become
\begin{eqnarray}\label{equ:i2}
\left[\begin{matrix}Y_1\\ \vdots\\ Y_L\end{matrix}\right]_{Lm\times 1}=\left[\begin{matrix}\hat{\mathbf{S}}_1^j\\ \vdots\\ \hat{\mathbf{S}}_L^j\end{matrix}\right]_{Lm\times n}\left[\begin{matrix}\sqrt{\hat{p}_1^j}\\ \vdots\\ \sqrt{\hat{p}_n^j}\end{matrix}\right]_{n\times 1}
\end{eqnarray}

We should of course guarantee two main postulates in the above procedure. Firstly, we should make sure that the set of linear equations in (\ref{equ:i2}) has a unique least square solution or in other words the matrix $\hat{\mathbf{S}}^j$ has rank $n$. Secondly, it should be proved that the above procedure will converge to the actual value of user powers. In the following, we express conditions in which the convergence of iterative method is guaranteed.

\begin{theo}\label{th3}
Suppose $\mathbf{S}$ contains a row that has no zero and $\hat{X}_i$'s are uniformly chosen from $\{\pm1\}^n$. Then the probability that $\hat{\mathbf{S}}_{Lm\times n}=[\hat{\mathbf{S}}_1^T\ldots\hat{\mathbf{S}}_L^T]^T$ has rank $n$ tends to one as $L$ approaches infinity, where $\hat{\mathbf{S}}_i$ is formed by multiplying each column of the signature matrix $\mathbf{S}_{m\times n}$ by the corresponding element of $\hat{X}_i$. 
\end{theo}

\emph{Proof}: Suppose that the $q^{th}$ row of $\mathbf{S}$ has no zero. Form an $L\times n$ matrix $\mathbf{C}$ whose $k^{th}$ row is the $(km+q)^{th}$ row of $\hat{\mathbf{S}}$ for $k=0,\dots,(L-1)$. Since the $\text{rank}(\hat{\mathbf{S}})\geq \text{rank}(\mathbf{C})$, it is sufficient to show that the $\text{rank}(\mathbf{C})=n$ with probability one for sufficiently large values of $L$. This matrix can be written as:
\begin{eqnarray}
\mathbf{C}=\left[\begin{matrix}X_1^T\\ \vdots\\ X_L^T\end{matrix}\right]\left[\begin{matrix}s_{q, 1}& &0\\ &\ddots&\\0& &s_{q, n}\end{matrix}\right]
\end{eqnarray}
where $s_{q, i}$ is the $i^{th}$ entry of the $q^{th}$ row of $\mathbf{S}$. Note that the second matrix is diagonal with rank $n$. Thus it suffices to show that the rank of the first matrix approches $n$ as $L$ increases. Let $P_L$ be the probabilty of having $n$ linearly independent vectors among $L$ $1\times n$ random vectors. By partitioning $\mathbf{C}$ into $\lfloor{L/n}\rfloor$ blocks we have the following inequality:
\begin{eqnarray}\label{equ:ireal}
P_L\geq 1-\left(1-P_n\right)^{\lfloor{L/n}\rfloor}
\end{eqnarray}
where $P_n$ is the probability of a random $n\times n$ matrix with $\{\pm1\}$ entries being invertible. The probability $P_n$ is lower bounded by \cite{Tao} 
\begin{eqnarray}
P_n\geq 1-\left(\frac{3}{4}+o(1)\right)^n.
\end{eqnarray}
Thus we have:
\begin{eqnarray}\label{equ:ibinary}
P_L\geq 1-\left(\frac{3}{4}+o(1)\right)^{n\lfloor{L/n}\rfloor}
\end{eqnarray}
According to (\ref{equ:ibinary}) $P_L$ tends to unity as $L$ approaches infinity and the proof is complete.$\blacksquare$
\\

According to \cite{Hossein}, for any uniqely decodable signature matrix there is an $\eta_{sup}$ such that if $p_i$'s belong to $\left[1-\eta_{sup},1+\eta_{sup}\right]$, then the user data can be extracted without any error. The following theorem shows that in this situation the user powers can be determind exactly with propability one.     
\begin{theo}\label{th4}
For any uniquely decodable signature matrix and the corresponding $\eta_{sup}$ explained in \cite{Hossein}, if $p_i$'s belong to $\left[1-\eta_{sup},1+\eta_{sup}\right]$, then in a noiseless channel we can exactly find user powers with probability one as $L$ approaches infinity.
\end{theo}
\emph{Proof}: According to \cite{Hossein}, in the first step of the above procedure, $\hat{X}_i^1$'s are exactly the user data which are random. Therefore the set of equations in (\ref{equ:i2}) are consistent, i.e. it has an exact solution. Moreover, Theorem \ref{th3} states that as $L$ increases, the probability of having a unique least square solution in (\ref{equ:i2}) approaches unity. Hence the resultant powers after the first step are the desired results.$\blacksquare$
\\

\begin{figure}[t]
\centering
\includegraphics[width=9cm]{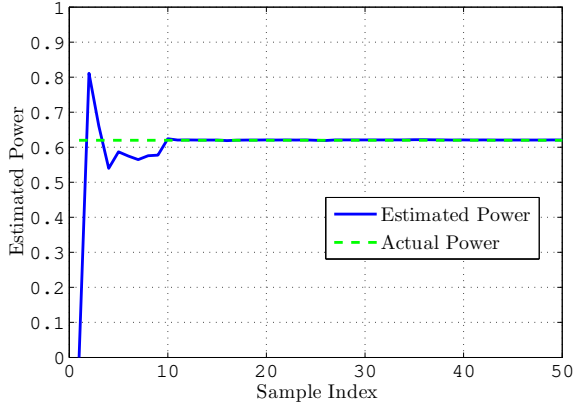}
\caption{Estimated user powers using the iterative method with $4$ iterations for a noiseless channel, $\eta=0.5$ .}
\label{fig:i1}
\end{figure}

In the following, we have simulated the above method. In the simulations, the signature matrix is an $8\times 13$ COW matrix proposed in \cite{813} and the initial value for the powers in the first iteration is unity. Fig. \ref{fig:i1} shows the result for a noiseless channel. Although according to \cite{Hossein}, the corresponding $\eta_{sup}$ is equal to $0.23$, we simulated the system for $\eta=0.5$. Notice that although Theorem \ref{th4} guarantees perfect estimation for $\eta<0.23$, Fig. \ref{fig:i1} depicts that even for higher fluctuations in the user powers, the performance of the method is impressive and the powers are derived after $4$ iterations.

\begin{figure}[t]
\centering
\includegraphics[width=9cm]{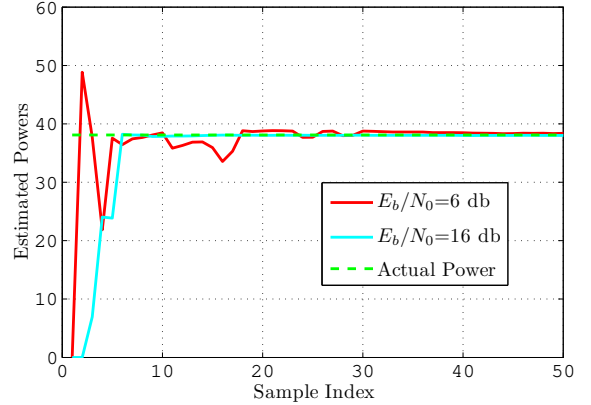}
\caption{Estimated user power using the iterative method with $4$ iterations. In this case user power is a Gaussian random variable with mean $25$ and standard deviation $10$. This simulation is done for two different values of   $E_b/N_0$, $6\text{dB}$ and  $16\text{dB}$.} 
\label{fig:i2}
\end{figure}

As previously stated, the constraints introduced in Theorem \ref{th4} is not the only situation that this method works. Simulation results prove that this method is not only useful for noisy channels but it is also applicable for high fluctuations of the user powers. It is expected that the effect of noise can be somehow compensated by finding the least square solution as shown in Fig. \ref{fig:i2}. In this case, the user power is a Gaussian random variable with mean $25$ and standard deviation $10$. This simulation is performed under two different values of $E_b/N_0$, $6\text{dB}$ and $16\text{dB}$. Interestingly, the powers are obtaind using only $30$ received vectors.

Simulation results show that the variance of the user powers is critical and their means have nothing to do with the estimaion procedure.

\begin{figure}[t]
\centering
\includegraphics[width=9cm]{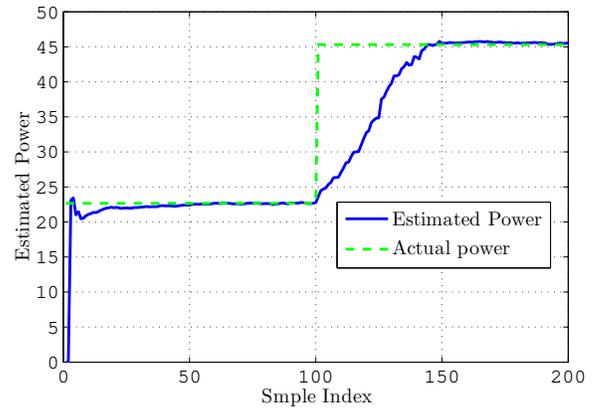}
\caption{Estimated user power using the iterative method with $4$ iterations. In this case user powers are doubled abruptly and $E_b/N_0=6\text{dB}$.} 
\label{fig:i3}
\end{figure}

\begin{figure}[t]
\centering
\includegraphics[width=9cm]{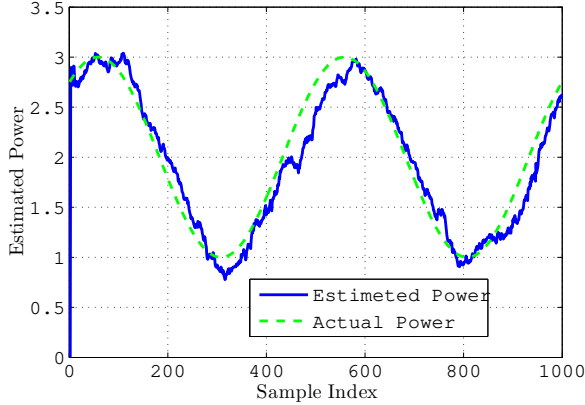}
\caption{Estimated user power using the iterative method with $4$ iterations. In this case user powers have sinusoidal shape and $E_b/N_0=6\text{dB}$.} 
\label{fig:i4}
\end{figure}

Figs. \ref{fig:i3} and \ref{fig:i4} show the ability of this method in tracking the user power changes. These curves are obtained with only $4$ iterations and 
$E_b/N_0=6dB$. We have used a window which keeps the last $40$ received vectors for estimation. As you can see this method estimate powers more rapidly and accurately in comparison with suboptimum method.     

\section{Bit-Error-Rate Performance}\label{sec:simulation}
In this section binary CDMA systems that use the suboptimum and iterative power estimations have been simulated. A $64\times 104$ COW matrix \cite{813} as the signature matrix is used. 

The simulation is performed for four different cases. In the first case, user powers are known at the receiver end, i.e., perfect power control. In other cases the performance of the system is simulated for suboptimum and iterative estimators and they are compared to the no power control case.

These simualations are performed for two different cases. In the first one, user powers are modeled as a sinusoidal changing variable in the range of $[0.5,1.5]$ with period of $20000$ samples. In the second case, user powers have piecewise constant variation with length $10000$ samples and abrupt changes to randomly chosen amplitudes from the $[0.5,1.5]$. In fact, these fluctuations for user powers are much faster than what appears in practice. 

In \cite{813}, a very fast ML decoding scheme has been proposed for COW codes. This method can be implemented with a similar number of computations for the case of the near-far effect. In order to make the decoder of \cite{813} suitable for the proposed model, it is necessary to make a minor adjustment. After estimating the powers $\hat{p}_1,\dots,\hat{p}_n$ the decoding algorithm has the following two steps:
\begin{itemize} 
\item Step 1: Let $\mathbf{S}_{m\times n}=[\mathbf{A}_{m\times m}\vline\mathbf{B}_{m\times(n-m)}]$ where $\mathbf{A}$ is an invertible matrix.
\item Step 2: Assume $\hat{X}=[\hat{X}^T_1~\hat{X}^T_2]^T\in\{\pm 1\}^n$ where $\hat{X}_1$ and $\hat{X}_2$ are $m\times 1$ and $(n-m)\times 1$, respectively. Find $\hat{X}_2$ that minimizes
\begin{eqnarray}
\|\mathbf{A}^{-1}Y&-&\mathbf{A}^{-1}\mathbf{B}\acute{\mathbf{P}}_2\hat{X}_2\nonumber\\&-&\acute{\mathbf{P}}_1\text{sign}(\mathbf{A}^{-1}Y-\mathbf{A}^{-1}\mathbf{B}\acute{\mathbf{P}}_2\hat{X}_2)\|
\end{eqnarray} 
and set
\begin{eqnarray}
\hat{X}_1=\text{sign}(\mathbf{A}^{-1}Y-\mathbf{A}^{-1}\mathbf{B}\acute{\mathbf{P}}_2\hat{X}_2)
\end{eqnarray}
where
\begin{eqnarray}
\acute{\mathbf{P}}_1=\text{diag}\left(p_{1},\dots,p_{m}\right)
\end{eqnarray}
and
\begin{eqnarray}
\acute{\mathbf{P}}_2=\text{diag}\left(p_{n-m+1},\dots,p_{n}\right).
\end{eqnarray}
\end{itemize}

\begin{figure}[t]
\centering
\includegraphics[width=9cm]{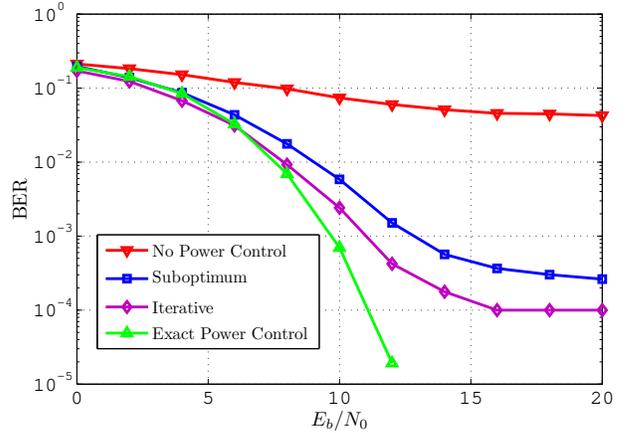}
\caption{Comparison of BER vs. $E_b/N_0$ for the BPSK modulation CDMA systems using the suboptimum and iterative power estimations when user powers have sinusoidal fluctuations; the extreme cases of perfect and no power controls are also depicted. This CDMA system consists of $64$ chips and $104$ users.}
\label{fig:s1}
\end{figure}

\begin{figure}[t]
\centering
\includegraphics[width=9cm]{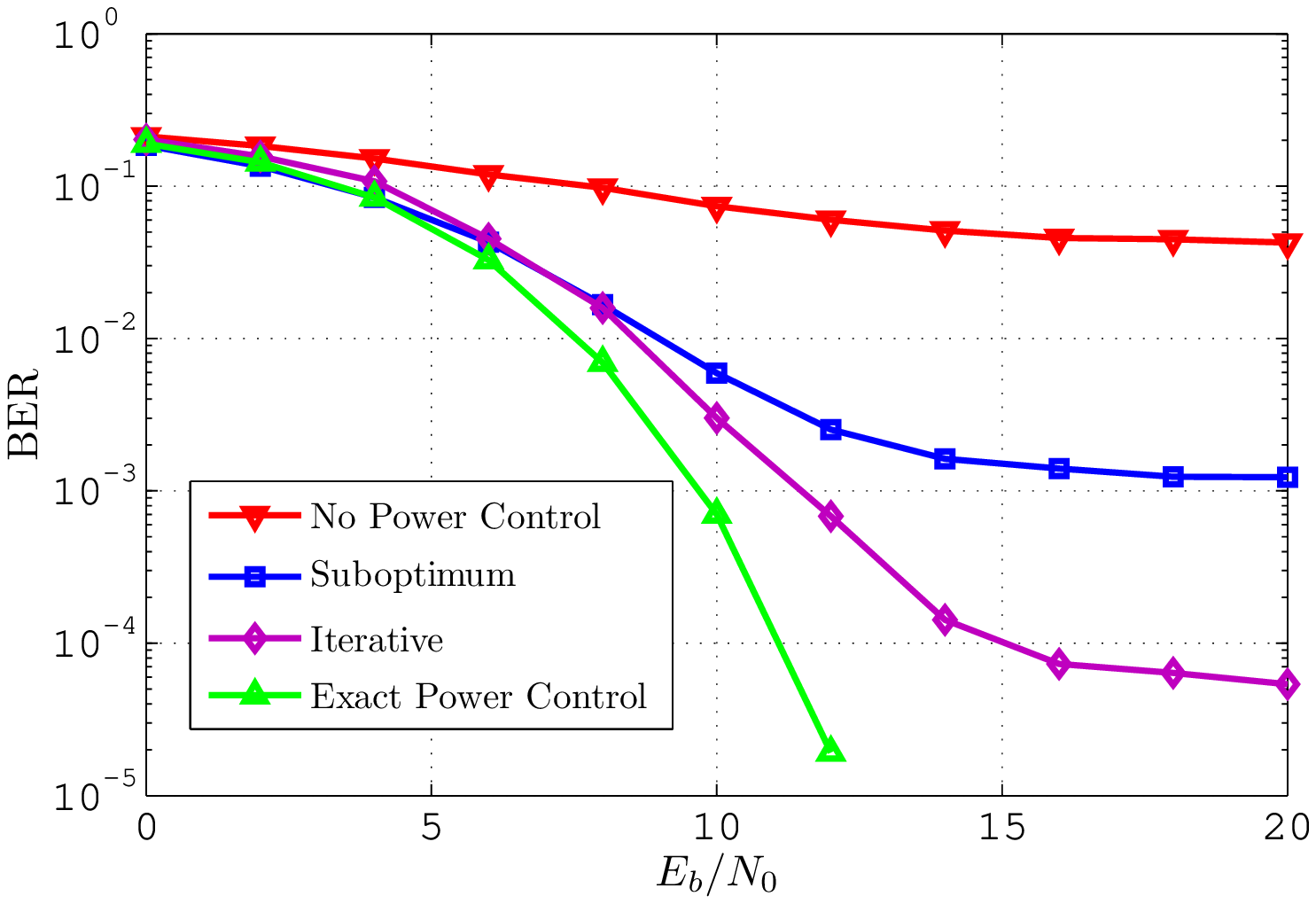}
\caption{Comparison of BER vs. Eb/N0 for the BPSK modulation CDMA systems using the suboptimum and iterative power estimations when user powers have stepwise fluctuations; the extreme cases of perfect and no power controls are also depicted. This CDMA system consists of $64$ chips and $104$ users.}
\label{fig:s2}
\end{figure}

Figs. \ref{fig:s1} and \ref{fig:s2} show the simulation results for sinusoidal and stepwise fluctuations, respectively. We use only $4$ iterations in these figures for the iterative method.

Note that both methods saturate for high values of $E_b/N_0$. This is because when the power changes, it takes some time for both methods to track the power variations. In this period, we always have some errors whether there is noise or not. 

\section{Conclusion and Future Studies}\label{sec:conclusion}
In this paper, the optimum power estimation (i.e. ML estimator) for a given model of stochastic process of user powers has been discussed. We have also introduced a lower complexity power estimation method which can perfectly estimate the user powers in a highly overloaded CDMA system. In fact simulation results have shown that the ML estimation and its simplified low complex version have similar performances. This paper also introduced a new characteristic for the signature matrices that shows their suitability for estimating the user powers. It was shown that for a system with $m$ chips, the user powers are not attainable by the ML method unless the number of users is less than $m(m+1)/2$. This characteristic is independent of the previously known characteristics such as WBE \cite{Massey} or COW \cite{813} and thus it should be incorporated seprately when designing the signature sequences for a CDMA system. While these methods do not use the knowledge of the input alphabet of the users, we have proposed an iterative method that works for binary input vectors and its performance is much better than the previous method.
 
Finding the optimum signature sequences according to the proposed characteristic for every chip rate $m$ and the number of users $n$ is an interesting topic for future research. By exploiting this knowledge, it would be possible to achive more efficient power estimation methods for binary input systems. The extension of this work to asynchronous CDMA systems would be another worthwhile research.

\section*{Acknowledgment}\label{sec:ack}
We would like to thank Ms P. Mansourifard for her useful comments.

\end{document}